\documentclass[10pt]{article}
\newcommand{\bea}{\begin{eqnarray}}
\newcommand{\eea}{\end{eqnarray}}
\newcommand{\beq}{\begin{equation}}
\newcommand{\eeq}{\end{equation}}
\newcommand{\nn}{\nonumber}
\newcommand{\aea}{&=&}
\newcommand{\apa}{&+&}

\newcommand{\Slash}[1]{\ooalign{\hfil/\hfil\crcr$#1$}}

\newcommand{\bra}{\langle}
\newcommand{\ket}{\rangle}

\newcommand{\braPSp}{\bra pS_{\perp}|}
\newcommand{\ketPSp}{|pS_{\perp} \ket}

\newcommand{\Sp}{S_{\perp}}

\newcommand{\hats}{\hat{s}}
\newcommand{\hatt}{\hat{t}}
\newcommand{\hatu}{\hat{u}}

\newcommand{\shs}{\hat{s}^2}
\newcommand{\ths}{\hat{t}^2}
\newcommand{\uhs}{\hat{u}^2}

\newcommand{\hsigma}{\hat{\sigma}}

\newcommand{\fsi}{\left(1+\frac{\hatu}{\hatt}\right)}
\newcommand{\dab}{\delta_{ab}}
\newcommand{\dac}{\delta_{ac}}
\newcommand{\dbc}{\delta_{bc}}

\newcommand{\Ncs}{N_c^2}

\newcommand{\SGPa}{G_{F}^{a}(x,x)-x\frac{dG_{F}^{a}(x,x)}{dx}}

\newcommand{\SFPa}{G_{F}^{a}(0,x)+\widetilde{G}_{F}^{a}(0,x)}

\usepackage{multicol}
\usepackage[dvips]{graphics}
\usepackage{amssymb}

\begin{document}

\begin{center}
\Large{New Analysis of the Single Transverse-Spin Asymmetry for
Hadron Production at RHIC} 

\vspace{0.5cm}

\normalsize{Koichi Kanazawa$^1$ and Yuji Koike$^2$\\
{

\vspace{0.3cm}

\it $^1$Graduate School of Science and Technology, Niigata
 University, Ikarashi, Niigata 950-2181, Japan\\
$^2$Department of Physics, Niigata University, Ikarashi, Niigata
 950-2181, Japan}}

\end{center}

\vspace{0.5cm}

\begin{center}
\textbf{Abstract}
\end{center}

\small{
We analyze the single transverse-spin
 asymmetry (SSAs) for inclusive hadron production in the $pp$ collision at RHIC
 based on the twist-3 mechanism in QCD.  As an origin of SSAs, we take into account of all kinds of pole
 contributions associated with the twist-3 quark-gluon
 correlation functions in the polarized proton.  
By the inclusion of the soft-fermion-pole contribution
in addition to the soft-gluon-pole contribution, we find SSAs observed at
RHIC are better described for all kinds of mesons.  
$P_T$-dependence of the asymmetry and the comparison with the FNAL data
are also presented.   
}
\normalsize{}


\newpage


\section{Introduction}
Since the first observation of large single-spin asymmetry (SSA) 
in $p^\uparrow p\to \pi X$
by the FNAL-E704 Collaboration\,\cite{E704}, SSA
has been one of the hot topics in
high energy spin physics.  
Large SSA has also been reported in 
semi-inclusive deep inelastic scattering (SIDIS), 
$ep^\uparrow \to ehX$ $(h=\pi,\ K$)\,\cite{COMPASS,HERMES}.
In the $pp$ collision, 
BNL-RHIC reported a similar large SSA
at even higher energies 
\cite{STAR1,STAR2,PHENIX,BRAHMS}.  
These large SSAs can not be explained within the conventional framework based on the  
parton model and perturbative QCD\,\cite{Kane} which had been successful 
in describing various hard inclusive processes.  
They reflect novel quark-gluon substructure of hadrons 
beyond the parton model picture. 

By now QCD 
mechanisms leading to such large SSA have been understood in some detail,
and they are often classified into two categories
depending on the kinematic regions they cover.   
One is the so-called ``T-odd" distribution and fragmentation functions
in the framework of the transverse momentum dependent (TMD) factorization.
This mechanism describes the SSAs in the region of the small
transverse momentum of the final hadron
as a result of a correlation between 
the spin and the intrinsic transverse momentum of
partons\,\cite{Sivers,Collins,Collins02,BJY03,BMP03,CM04,MM09}.    
Although this approach provides a physical
interpretation of SSA, proof for the TMD factorization  
has been limited to the almost-back-to-back
jets production in $e^+e^-$ annihilation\,\cite{CS81}, Drell-Yan\,\cite{CSS85} 
and SIDIS\,\cite{JMY05}.
In addition, the universality property of the TMD distribution function
becomes unclear, in particular, for hadron-hadron collisions\,\cite{BMP04}.   
From the phenomenological analyses of SIDIS data given by CERN and DESY, 
a parametrization of the responsible TMD functions has been obtained\,\cite{Anselmino}.

Another mechanism describes the SSA
as a twist-3 effect in the 
framework of the collinear factorization
\,\cite{ET,QS92,QS99,Kanazawa,EKT,Kouvaris,KT071,KT072,KQ08,KoikeTomita}.  
This mechanism is designed for the description of SSA 
in the region where the transverse momentum of the final hadron
can be regarded as a hard scale in the process.  
In this mechanism
twist-3 quark-gluon correlation functions replace the parton 
distribution and/or fragmentation functions,
and gives the information on the multiparton correlations in the hadron.  
Although the above two mechanisms describe SSA in different kinematic regions, 
recent studies have shown that for 
a certain component contributing to SSA they give the equivalent description for SSA
in the intermediate region of the transverse momentum for which
both frameworks are valid, and these constitute a unique and unified  
QCD description for SSA\,\cite{OL}.

In this paper we study SSA for the hadron ($\pi$, $K$) 
productions in the $pp$ collision at the RHIC energies ($\sqrt{S}=200$ and 62.4 GeV), 
\beq
p^{\uparrow}(p,\Sp) + p(p') \to h(P_h) + X,
\label{process}
\eeq
in the framework of the collinear factorization, extending the previous 
works\,\cite{QS99,Kanazawa,Kouvaris,KoikeTomita}.  
Here $\Sp$ represents the spin vector for the 
transversely polarized nucleon, and $p,\,p',\,P_h$ are the 4-momenta of each particle. 
The SSA for the above process is defined as $A_N= (\sigma^\uparrow-\sigma^\downarrow) /
(\sigma^\uparrow+\sigma^\downarrow)$ where
$\sigma^{\uparrow(\downarrow)}$ are the cross section for the process (\ref{process}) 
corresponding to $\Sp$ ($-\Sp$). 
In this framework, three kinds of twist-3 cross section
contribute to the single-spin-dependent cross section 
$\Delta\sigma\equiv\sigma^\uparrow-\sigma^\downarrow$: 
\bea
\Delta\sigma &=& G^{(3)}(x_1,x_2) \otimes f(x') \otimes D(z) \otimes
\hat{\sigma}_A \nn\\
\apa h(x) \otimes f(x') \otimes \widehat{E}^{(3)}(z_1,z_2) \otimes
\hat{\sigma}_B \nn\\
\apa h(x) \otimes E^{(3)}(x_1',x_2') \otimes D(z) \otimes
\hat{\sigma}_C, 
\label{3tu}
\eea
where $\otimes$
denotes convolution with respect to the light-cone momentum fraction, and
$f$, $h$, $D$ are, respectively, twist-2 unpolarized distribution, transversity
distribution and the fragmentation function.   Two
variable functions with the upper index $(3)$ are the twist-3 multiparton correlation
functions:  $G^{(3)}(x_1,x_2)$, $E^{(3)}(x_1',x_2')$ and $\widehat{E}^{(3)}(z_1,z_2)$ are,
respectively, the distribution function in the polarized nucleon, distribution function
in the unpolarized nucleon and the twist-3 fragmentation function for the final hadron.  
$\hat{\sigma}_A,\hat{\sigma}_B,\hat{\sigma}_C$ are the perturbatively calculable 
partonic hard cross sections.  
It has been shown in \cite{Kanazawa} that
the third term in (\ref{3tu}) can be neglected compared to the
first term because of the smallness of the partonic cross section; 
$\hat{\sigma}_C \ll \hat{\sigma}_A$.   
The second term of
(\ref{3tu}) involves the twist-3 fragmentation function, 
for which one needs different formalism to calculate the cross section\,\cite{YZ09}.   
In this circumstance, phenomenological analysis of the data 
has been performed including only the first term of (\ref{3tu})\,\cite{QS99,Kouvaris}.
According to the twist-3
formalism for SSA coming from the twist-3 distribution, 
the cross section occurs from a pole part of
an internal propagator in the hard part (See e.g. \cite{QS92,EKT}
for the detail of the formalism).  
These poles fix one of the momentum fractions in the twist-3 
distribution, and those poles are classified into the soft-gluon-pole (SGP) and
the soft-fermion-pole (SFP) corresponding, respectively, to
the zero momentum of the gluonline and the quarkline in the
twist-3 distribution.  
A characteristic feature 
of the SGP contribution is that some of them 
accompany a ``derivative" of the SGP
function, which enhances the cross section compared with the usual ``nonderivative" terms.  
Accordingly it has been considered to be a main source of SSA compared with
the SFP contribution, and the previous analyses of $A_N$ data 
focussed on the former\,\cite{QS99,Kouvaris}.  
Even though the SFP contribution does not appear with the derivative,
there is no clue that the corresponding nonperturbative function itself
is small compared with the SGP function and therefore
its importance depends on the magnitude of the partonic hard cross
section.  In a recent paper\,\cite{KoikeTomita}, one of the authors
has shown that the partonic hard cross section for the SFP contribution
is much larger than those for the SGP contribution, and the total
SFP contribution could be a large source of SSA if the SFP function
has a comparable magnitude as the SGP function.

The purpose of this paper is to present a numerical analysis of 
$A_N$ for the light-hadron production at RHIC, $p^\uparrow p\to h X$ ($h=\pi$, $K$), 
including all the effects from the twist-3 quark-gluon correlation functions
contributing to the first term of (\ref{3tu}).  
In principle, independent twist-3 quark-gluon distribution functions
should be determined by performing a global analysis of various SSA data
for SIDIS and the $pp$ collisions.  
However, SSA data obtained at HERMES and COMPASS are in the low $P_T$ region,
and therefore is not suited for the analysis by the twist-3 formalism.
We thus focus on the asymmetry in the $pp$ collision in this analysis.  
Previous analysis of $A_N$ in the $pp$ collision included only the SGP contribution\,\cite{QS99,Kouvaris}, 
and 
could reproduce the $A_N$ obtained at FNAL and 
RHIC reasonably well, albeit with a deviation for $A_N^{K^-}$
and the $P_T$ dependence\,\cite{Kouvaris,STAR2}. 
We will see in this paper that the SGP contribution itself is insufficient
to understand the patterns shown in the RHIC $A_N$ data for the $\pi$ and $K$ production and
inclusion of the SFP contribution significantly improves the agreement with the data.  
This suggests that the SFP contribution could be an important candidate for the origin of
SSA.  

The rest of the paper is organized as follows:  In Sec 2, we
summarize the necessary cross section formula as well as the twist-3
quark-gluon correlation functions.  In Sec 3, we present our
method of numerical analysis and the result.  Section 4 is devoted to a brief summary.  

\section{Cross-section formula}
 
There are two independent
twist-3 quark-gluon correlation functions, $G_F$ and $\widetilde{G}_F$, 
contributing to the first term of
(\ref{3tu}).  They are defined from the nucleon matrix element 
of the composite quark-gluon operator on the light-cone (See \cite{EKT,KoikeTomita} for the detail): 
\bea
&& \int \frac{d\lambda}{2\pi} \int
\frac{d\mu}{2\pi} e^{i\lambda x_1}
e^{i\mu(x_2-x_1)} \braPSp \bar{\psi}^a_j(0) g F^{\alpha\beta}(\mu n)
n_{\beta} \psi^a_i(\lambda n) \ketPSp \nn\\
&& \qquad\qquad= \frac{M_N}{4}(\Slash{p})_{ij} \epsilon^{\alpha pn\Sp} G_{F}^a(x_1,x_2) +
i\frac{M_N}{4} (\gamma_5\Slash{p})_{ij} \Sp^{\alpha}
\widetilde{G}_F^a(x_1,x_2) + \cdots .
\label{MF}
\eea
where $n$ is a lightlike vector satisfying $p\cdot n=1$, 
$\psi^a_i$ is the quark field for flavor $a$ with the spinor index $i$, 
$F^{\alpha\beta}$ is the gluon's field strength, 
$\epsilon^{\alpha pn\Sp} \equiv\epsilon^{\alpha\mu\nu\rho}
p_{\mu}n_{\nu} S_{\perp\rho}$ ($\epsilon_{0123}\equiv +1$), and $\cdots$ denotes twist-4 or higher.  
The nucleon mass $M_N$ is introduced to define 
$G_F$ and $\widetilde{G}_F$ dimensionless. 
From parity and $T$ invariance, 
$G_F^a(x_1,x_2)$ and $\widetilde{G}^a_F(x_1,x_2)$
satisfy the relation 
\bea
G_F^a(x_1,x_2)=G_F^a(x_2,x_1),\quad
\widetilde{G}_F^a(x_1,x_2)=-\widetilde{G}_F^a(x_2,x_1).
\label{symmetry}
\eea
The twist-3 correlation function for antiquark flavor can be defined similarly
to (\ref{MF}), replacing the nonlocal operator
in (\ref{MF}),
$\bar{\psi}^a_j(0) g F^{\alpha\beta}(\mu n)
n_{\beta} \psi^a_i(\lambda n)$, by its charge conjugated operator.
The ``antiquark" twist-3 correlation function obtained this way is related to the original
``quark" correlation function as
\bea
G_F^{\bar{a}}(x_1,x_2) = G_F^{a}(-x_2,-x_1),\quad
\widetilde{G}_F^{\bar{a}}(x_1,x_2)=-\widetilde{G}_F^a(-x_2,-x_1). 
\label{anti}
\eea
The contribution to the single-spin-dependent
cross section from the correlation function (\ref{MF})
has been obtained in the literature. 
For completeness we give the explicit formula here in our notation:  
The SGP contribution is given by\,\cite{Kouvaris}
\bea
P^0_h\frac{d^3\Delta\sigma^{SGP}}{dP_h^3} \aea \frac{\pi M_N\alpha_S^2}{S}
\epsilon^{pnP_h\Sp} \int^1_{z_{min}} \frac{dz}{z^3} \int^1_{x'_{min}}
\frac{dx'}{x'} \int \frac{dx}{x} \frac{1}{x'S+T/z} \delta \left(
x-\frac{-x'U/z}{x'S+T/z} \right) \nn\\
&\times& \left(\frac{1}{-\hatu} \right) \biggl[ \sum_{a,b,c}
\left( \SGPa \right) \times \nn\\
&& \{ f_b(x') (D_c(z) \hsigma^{G}_{ab\to c} + D_{\bar{c}}(z)\hsigma^{G}_{ab\to
\bar{c}})
+f_{\bar{b}}(x') (D_c(z)\hsigma^{G}_{a\bar{b}\to
 c}+D_{\bar{c}}(z)\hsigma^{G}_{a\bar{b}\to\bar{c}}) \} \nn\\
&+& \sum_a\left( \SGPa \right) f_{\bar{a}}(x') D_g(z)
\hsigma^{G}_{a\bar{a}\to g}\nn\\
&+& \sum_a\left( \SGPa \right) G(x') (D_a(z) \hsigma^{G}_{ag\to a} +
D_g(z)\hsigma^{G}_{ag\to g}) \biggr],
\label{tSGP}
\eea
where $\alpha_s=g^2/(4\pi)$ is the strong coupling constant, 
the Mandelstam variables for the process are defined as
\bea
S \aea (p+p')^2, \nn\\
T \aea (p-P_h)^2, \nn\\
U \aea (p'-P_h)^2, \nn
\eea
and the lower limits of the integration are given by
$z_{min}=-\frac{T+U}{S},x'_{min}=\frac{-T/z}{S+U/z}$.
$f_a(x)$ and $D_a(z)$ are, respectively, distribution and fragmentation functions
for quark (or antiquark) flavor $a$.  
$G(x)$ and $D_g(z)$ are, respectively, distribution and fragmentation functions
for gluon.  
Partonic hard cross section in (\ref{tSGP}) are expressed in terms of the
Mandelstam variables in the parton level:
\bea
\hats \aea (xp+x'p')^2, \nn\\
\hatt \aea (xp-P_h/z)^2, \nn\\
\hatu \aea (x'p'-P_h/z)^2, \nn
\eea
and they are given by 
\bea
%
%
\hsigma^{G}_{ab\to c} \aea \left[ \frac{1}{N_c^2} + \frac{1}{2N_c^2}\fsi
\right] \left(\frac{\hats^2+\hatu^2}{\hatt^2}\right) \dac \nn\\
\apa \left[ \frac{1}{N_c^2} - \frac{N_c^2-2}{2N_c^2}\fsi \right]
\left(\frac{\hats^2+\hatt^2}{\hatu^2}\right) \dbc \nn\\
\apa \left[ -\frac{N_c^2+1}{N_c^3} - \frac{1}{N_c^3} \fsi \right]
\left(\frac{\hats^2}{\hatt\hatu}\right) \dac \dbc, \\
%
%
\hsigma^{G}_{ab \to \bar{c}} \aea 0, \\
%
%
\hsigma^{G}_{a\bar{b}\to c} \aea \left[ \frac{N_c^2-2}{2N_c^2} +
\frac{1}{2N_c^2} \fsi \right]
\left(\frac{\hats^2+\hatu^2}{\hatt^2}\right) \dac \nn\\
\apa \left[ -\frac{1}{2N_c^2} - \frac{N_c^2-2}{2N_c^2} \fsi \right]
\left(\frac{\hatt^2+\hatu^2}{\hats^2}\right) \dab \nn\\
\apa \left[ -\frac{1}{N_c^3} - \frac{1}{N_c^3} \fsi \right]
\left(\frac{\hatu^2}{\hats\hatt}\right) \dac \dbc, \\
%
%
\hsigma^{G}_{a\bar{b}\to \bar{c}} \aea \left[ \frac{N_c^2-2}{2N_c^2} -
\frac{1}{N_c^2}\fsi \right] \left(\frac{\hats^2+\hatt^2}{\hatu^2}\right)
\dbc \nn\\
\apa \left[ -\frac{1}{2N_c^2} - \frac{1}{N_c^2} \fsi \right]
\left(\frac{\hatt^2+\hatu^2}{\hats^2}\right) \dab \nn\\
\apa \left[ -\frac{1}{N_c^3} - \frac{1}{N_c^3} \fsi \right]
\left(\frac{\hatt^2}{\hats\hatu}\right) \dac \dbc, \\
%
%
\hsigma^{G}_{a\bar{a}\to g} \aea \frac{1}{2N_c^3}
\left(\frac{\hatt}{\hatu}+\frac{\hatu}{\hatt}\right)
\left(1+2N_c^2\frac{\hatt\hatu}{\hats^2}\right) \nn\\
\apa \frac{1}{2N_c} \fsi \left(\frac{\hatt}{\hatu}+\frac{\hatu}{\hatt}\right)
\left(1-N_c^2\frac{\hatu^2}{\hats^2}\right), \\
%
%
\hsigma^{G}_{ag\to a} \aea \frac{1}{2\left(N_c^2-1\right)}
\left(\frac{\hats}{\hatu}+\frac{\hatu}{\hats}\right)
\left(1-N_c^2\frac{\hatu^2}{\hatt^2}\right) \nn\\
\apa \frac{1}{2N_c^2\left(N_c^2-1\right)} \fsi
\left(\frac{\hats}{\hatu}+\frac{\hatu}{\hats}\right)
\left(1+2N_c^2\frac{\hats\hatu}{\hatt^2}\right), \\
%
%
\hsigma^{G}_{ag\to g} \aea \frac{1}{2\left(N_c^2-1\right)}
\left(\frac{\hats}{\hatt}+\frac{\hatt}{\hats}\right)
\left(1-N_c^2\frac{\hatt^2}{\hatu^2}\right) \nn\\
\apa \frac{-1}{2\left(N_c^2-1\right)} \fsi
\left(\frac{\hats}{\hatt}+\frac{\hatt}{\hats}\right)
\left(1-N_c^2\frac{\hats^2}{\hatu^2}\right),
\eea
where $N_c=3$ is the number of colors for SU(3).  
In (\ref{tSGP}), $\widetilde{G}_F^a$ does not contribute to the SGP contribution
because $\widetilde{G}_F^a(x,x)=0$ 
due to the symmetry property (\ref{symmetry}),
and $G_F^a(x,x)$ appears in the form of ${d G_F^a(x,x) \over dx}-G_F^a(x,x)$\,\cite{Kouvaris,KT072}.
The SFP contribution was obtained in \cite{KoikeTomita} as
\bea
P_h^0\frac{d^3\Delta\sigma^{SFP}}{dP_h^3} &=& \frac{\alpha_S^2}{S}
\frac{M_N\pi}{2} \epsilon^{pnP_h\Sp} \int^1_{z_{min}} \frac{dz}{z^3}
\int^1_{x'_{min}} \frac{dx'}{x'} \int \frac{dx}{x} \frac{1}{x'S+T/z}
\delta \left(x-\frac{-x'U/z}{x'S+T/z} \right) \nn\\
&\times& \biggl[ \sum_{a,b,c} \left(\SFPa\right) \times \nn\\ 
&& \{ f_b(x') (D_c(z)\hsigma^F_{ab\to c} + D_{\bar{c}}(z)\hsigma^F_{ab\to\bar{c}})
+ f_{\bar{b}}(x')
(D_c(z)\hsigma^F_{a\bar{b}\to c}+D_{\bar{c}}(z)\hsigma^F_{a\bar{b}\to\bar{c}})
\} \nn\\
\apa \sum_{a,b} \left(\SFPa\right) (f_b(x')D_g(z) \hsigma^F_{ab\to
g}+f_{\bar{b}}(x')D_g(z)\hsigma^F_{a\bar{b}\to g}) \nn\\
\apa \sum_{a,c} \left(\SFPa\right) G(x') (D_c(z)\hsigma^F_{ag\to
c}+D_{\bar{c}}(z)\hsigma^F_{ag\to \bar{c}}) \nn\\
\apa \sum_a \left(\SFPa\right) G(x') D_g(z) \hsigma^F_{ag\to g}
\biggr] \label{tSFP},
\eea
where the partonic hard cross sections are given by
\bea
%
\hsigma^F_{ab\to c} \aea
\frac{-(N_c^2\hats+2\hatt)(\hats^2+\hatu^2)}{N_c^2\hatt^3\hatu} \dac
+ \frac{-(N_c^2\hatt+\hatu-\hats)\hats}{N_c^3\hatt\hatu^2} \dab \dac, \\
%
%
\hsigma^F_{ab\to \bar{c}} \aea 0, \\
%
%
\hsigma^F_{a\bar{b}\to c} \aea
\frac{(N_c^2\hatu+2\hatt)(\hats^2+\hatu^2)}{N_c^2\hatt^3\hatu} \dac 
+ \frac{(N_c^2\hatu+2\hats)(\hatt^2+\hatu^2)}{N_c^2\hats^2\hatt\hatu} \dab
- \frac{(N_c^2-1)\hatu^2}{N_c^3\hats\hatt^2} \dab \dac, \\
%
%
\hsigma^F_{a\bar{b}\to \bar{c}} \aea
\frac{-(\Ncs\hatt+2\hats)(\ths+\uhs)}{\Ncs\shs\hatt\hatu} \dab 
+ \frac{-\Ncs\hats+\hatt-\hatu}{N_c^3\uhs} \dab \dac, \\
%
%
\hsigma^F_{ab\to g} \aea
\frac{(\Ncs\hats+2\hatt)(\shs+\uhs)}{\Ncs\hatt^3\hatu}
+ \frac{-1}{N_c^3\hats\hatt\uhs}
(\Ncs(\hats^3+3\shs\hatu-2\hatu^3)+\hats^3-\hats^2\hatu) \dab, \\
%
%
\hsigma^F_{a\bar{b}\to g} \aea
\frac{-(\Ncs\hatu+2\hatt)(\shs+\uhs)}{\Ncs\hatt^3\hatu} \nn\\
\apa \left[ \frac{1}{N_c^3}\left(\frac{\hatu}{\hats\hatt}+\frac{1}{\hatu}\right) +
\frac{1}{N_c}
\left(\frac{\shs+\hats\hatt+\ths}{\hats\uhs}-\frac{\hatu}{\ths}\right) +
\frac{N_c(\hatu^3-\hatt^3)(\ths+\uhs)}{\shs\ths\uhs} \right] \dab, \\
%
%
\hsigma^F_{ag\to c} \aea \left[
\frac{N_c^2(\hats^3-\hatu^3)(\shs+\uhs)}{(\Ncs-1)\hats\hatt^3\uhs}
+
\frac{\hats\hatu(\shs+\hats\hatu-\uhs)-\Ncs( \hats^4 + \hats^3\hatu + \shs\uhs
+ \hats\hatu^3 + \hatu^4 )}{\Ncs(\Ncs-1)\hats\ths\uhs}
\right] \dac \nn\\
\apa \frac{(\Ncs\hatu+2\hats)(\ths+\uhs)}{N_c(\Ncs-1)\shs\hatt\hatu}, \\
%
%
\hsigma^F_{ag\to \bar{c}} \aea
\frac{\hats+2\hatt-\Ncs\hats}{\Ncs(\Ncs-1)\uhs} \dac
+ \frac{-(\Ncs\hatt+2\hats)(\ths+\uhs)}{N_c(\Ncs-1)\shs\hatt\hatu}, \\
%
%
\hsigma^F_{ag\to g} \aea \frac{-\Ncs}{(\Ncs-1)\shs\hatt^3\uhs}
( 4\hats^6 + 11\hats^5\hatt + 19\hats^4\ths + 22 \hats^3\hatt^3 +
19\shs\hatt^4 + 11\hats \hatt^5 + 4\hatt^6 ) \nn\\
\apa \frac{1}{\Ncs(\Ncs-1)\hats\ths\uhs} ( - \hats\hatt\uhs +
\Ncs( \hats^4 + \hats^3\hatt + 2\shs\ths + \hats\hatt^3 + \hatt^4 ) ) .
\eea
The summation $\sum_{a,b,c}$, $\sum_{a,b}$, $\sum_a$ in (\ref{tSGP}) and (\ref{tSFP})
indicate that
the sum of $a$ is over all quark and antiquark flavors, and
the sum of $b$ and $c$ is over all quark flavors for the quark flavor $a$ and
is over all antiquark flavors for the antiquark flavor $a$.
As is shown in (\ref{tSFP}), the SFP contribution appears in the form of
$\SFPa$, which allows us to parametrize $\SFPa$ when we study the process
$p^{\uparrow}p\to hX$.

For completeness we also list the twist-2 unpolarized cross section for
$pp\to hX$\,\cite{unpol} which becomes the denominator of $A_N$: 
\bea
P_h^0 \frac{d^3\sigma^{unpol}}{dP_h^3} &=& \frac{\alpha_S^2}{S}
\int^1_{z_{min}} \frac{dz}{z^2} \int^1_{x'_{min}}
\frac{dx'}{x'} \int \frac{dx}{x} \frac{1}{x'S+T/z} \delta \left(
x-\frac{-x'U/z}{x'S+T/z} \right) \nn\\
&\times& \biggl[ \sum_{a,b,c} 
\{ f_a(x) f_b(x') ( D_{c }(z)\hsigma^U_{ab\to
c}+D_{\bar{c} }(z) \hsigma^U_{ab\to \bar{c}} ) \nn\\
&& + f_a(x) f_{\bar{b}}(x')
( D_{c }(z)\hsigma^U_{a\bar{b}\to
c}+D_{\bar{c} } (z)\hsigma^U_{a\bar{b}\to \bar{c}} )\} \nn\\
&+& \sum_a f_a(x) f_{\bar{a}}(x') D_{g }(z)
\hsigma^U_{a\bar{a}\to g}\nn\\
&+& \sum_a f_a(x) G(x') (D_{a}(z)\hsigma^U_{ag\to
a}+D_{g}(z)\hsigma^U_{ag\to g}) \nn\\
&+& \sum_a G(x) f_a(x') (D_a(z)\hsigma^U_{ga\to
a}+D_g(z)\hsigma^U_{ga\to g}) \nn\\
&+& \sum_a G(x) G(x') (D_{g}(z)\hsigma^U_{gg\to g}+D_{a}(z)\hsigma^U_{gg\to a})
 \biggl],
\eea
\bea
%
\hsigma^U_{ab\to c} \aea \frac{C_F}{N_c} \frac{\hats^2+\hatu^2}{\hatt^2}
\delta_{ac}
 + \frac{C_F}{N_c} \frac{\hats^2+\hatt^2}{\hatu^2}
\delta_{bc}
 - \frac{2C_F}{N_c^2} \frac{\hats^2}{\hatt\hatu}
\delta_{ac}\delta_{bc}, \\
%
\hsigma^U_{ab\to \bar{c}} \aea 0 , \\
%
%
\hsigma^U_{a\bar{b}\to c} \aea \frac{C_F}{N_c}
\frac{\hatt^2+\hatu^2}{\hats^2} \delta_{ab} 
+ \frac{C_F}{N_c} \frac{\hats^2+\hatu^2}{\hatt^2}\delta_{ac}
- \frac{2C_F}{N_c^2} \frac{\hatu^2}{\hats\hatt} \delta_{ac}\delta_{bc}, \\
%
%
\hsigma^U_{a\bar{b}\to \bar{c}} 
\aea \frac{C_F}{N_c} \frac{\hatt^2+\hatu^2}{\hats^2} \delta_{ab}
+ \frac{C_F}{N_c} \frac{\hats^2+\hatt^2}{\hatu^2} \delta_{bc}
- \frac{2C_F}{N_c^2} \frac{\hatt^2}{\hats\hatu} \delta_{ac}\delta_{bc}, \\
%
%
\hsigma^U_{a\bar{a}\to g} \aea \frac{2C_F^2}{N_c}
\frac{\hatu^2+\hatt^2}{\hatu\hatt}
- 2 C_F \frac{\hatu^2+\hatt^2}{\hats^2}, \\
%
%
\hsigma^U_{ag\to a} \aea -\frac{C_F}{N_c}
\frac{\hatu^2+\hats^2}{\hatu\hats}
+ \frac{\hatu^2+\hats^2}{\hatt^2}, \\
%
%
\hsigma^U_{ag\to g} \aea -\frac{C_F}{N_c}
\frac{\hats^2+\hatt^2}{\hats\hatt} 
+ \frac{\hats^2+\hatt^2}{\hatu^2}, \\
%
\hsigma^U_{ga\to a} \aea \hsigma^U_{ag\to g}\\
\hsigma^U_{ga\to g} \aea \hsigma^U_{ag\to a}\\
%
\hsigma^U_{gg\to a} \aea \frac{1}{2N_c} \frac{\hatu^2+\hatt^2}{\hatu\hatt}
- \frac{1}{2C_F} \frac{\hatu^2+\hatt^2}{\hats^2}, \\
%
%
\hsigma^U_{gg\to g} \aea \frac{9}{2} \left( 3-\frac{\hatu\hatt}{\hats^2} -
\frac{\hatu\hats}{\hatt^2} - \frac{\hats\hatt}{\hatu^2} \right),
\eea
where $C_F=(N_c^2-1)/(2N_c)$ with $N_c=3$.  
With the above defined cross section, $A_N$ for $p^\uparrow p\to hX$ ($h=\pi, K$) is given by 
\bea
A_N = \frac{
P_h^0\frac{d^3\Delta\sigma^{SGP}}{dP_h^3}
 + P_h^0\frac{d^3\Delta\sigma^{SFP}}{dP_h^3} }
{ P_h^0\frac{d^3\sigma^{unpol}}{dP_h^3} }.
\eea

\section{Numerical analysis}

Independent unknown functions appearing in $A_N$ 
are the SGP functions
\bea
G_{F}^a(x,x), 
\eea
and the SFP functions
\bea
\SFPa \label{SFPa}, 
\eea
for each quark and antiquark flavor ($a=u,\,d,\, s,\, \bar{u},\, \bar{d},\, \bar{s}$).   
Until now, the $A_N$ data obtained at RHIC are the only
ones to which twist-3 mechanism in the collinear factorization can be directly applied.
In fact, the next-to-leading-order (NLO) QCD can describe the unpolarized cross section
for the $pp\to\pi X$ process at RHIC energy, which constitutes
the denominator of $A_N$.  
We thus use the $A_N$ data by the STAR Collaboration at $\sqrt{S}=200$ GeV\,\cite{STAR1,STAR2}
and those by the BRAHMS Collaboration at $\sqrt{S}=62.4$ GeV\,\cite{BRAHMS}.  
All data for $A_N^{\pi^0}$ from the STAR Collaboration are in the range
of $P_{hT}\geq 1$ GeV, so we adopt all of the data presented in \cite{STAR1,STAR2} in the fitting.
On the other hand, some of the BRAHMS data are in the low $P_{hT}$ region.
But they reported data for $\pi^{\pm}$ and $K^\pm$, which are
important for disentangling the flavor structure of the
quark-gluon correlations in the nucleon. 
So we will adopt data for $\pi^\pm$ and $K^\pm$ in \cite{BRAHMS} 
with $P_{hT}\geq 0.7$ GeV in the fitting 
for this initial study on the SFP contribution.  
The total number of data points turn out to be 61.  
Unlike the analysis in \cite{Kouvaris}, we shall not use FNAL-E704 data in the fitting, 
since the NLO QCD can not reproduce the unpolarized cross section
at the fixed target energy\,\cite{hatan}. 

For the parametrization of the SGP and SFP functions, 
we assume the following form: 
\bea
& & G_{F}^{a}(x,x) = N_a^G x^{\alpha_a^G} (1-x)^{\beta_a^G} f_a(x),
\label{SGPF} \\
& & \SFPa = N_a^F x^{\alpha_a^F} (1-x)^{\beta_a^F} f_a(x),
\label{SFPF}
\eea
where $f_a(x)$ is the unpolarized quark distribution, and
$N_a^{G},\alpha_a^{G},\beta_a^{G}$, $N_a^{F},\alpha_a^{F},\beta_a^{F}$ are 
the parameters to be determined 
by the $\chi^2$ fitting to the data.  Of course, 
the twist-3 quark-gluon correlation functions
$G_F(x_1,x_2,\mu)$ and $\widetilde{G}_F(x_1,x_2,\mu)$ have their own
logarithmic scale dependence
($\mu$ dependence) which differs from that of the twist-2 unpolarized parton
densities\footnote{For the complete LO evolution equation for $G_F(x_1,x_2,\mu)$ and 
$\widetilde{G}_F(x_1,x_2,\mu)$, see \cite{Braun}.}
and therefore
the above ansatz for the SGP and SFP functions is not correct in the rigorous sense. 
However, 
at the present stage the SSA data is not so accurate, as the
precise scale dependence of the twist-3 correlation is required.  
We therefore assume that the SGP and SFP component of the twist-3 correlation functions
obey the same scale dependence as the twist-2 distribution
through the ansatz (\ref{SGPF}) and (\ref{SFPF}) for simplicity. 
For the unpolarized parton densities in the nucleon, we use the LO parametrization 
of GRV98\,\cite{GRV98}.
As for the fragmentation functions for pions and kaons,
we use the parametrization in \cite{DSS07} (hereafter referred to as the
DSS fragmentation function), which can describe a greater variety of data
in wider kinematic ranges, in particular for kaons, compared with the
older more common
ones in the literature.
When we calculate $A_N$
we set $\mu=P_{hT}$ for the STAR data and 
$\mu=1$ GeV for the BRAHMS data
in all distribution and fragmentation functions.  
In the $\chi^2$-fitting analysis below, we have used MINUIT package program in the 
CERN library.

Since
our main purpose here is to investigate the role and importance of the
SFP contribution in comparison to the SGP contribution,
we will analyze the data in terms of the following three
sets of the fitting functions:

\vspace{0.3cm}

\noindent
FIT 1: SGP + SFP contributions with 13 free parameters: 
\bea
&&N_u^G,\quad N_d^G,\quad N_s^G=N_{\bar{s}}^G,\quad N_{\bar{u}}^G=N_{\bar{d}}^G,\nonumber\\
&&\alpha^G_{u,d,s,\bar{u},\bar{d},\bar{s}}\equiv\alpha^G,\quad \beta^G_{u,d}\equiv\beta^G_v,\quad
\beta^G_{\bar{u},\bar{d},s,\bar{s}}\equiv\beta^G_s\nonumber\\
&&N_u^F,\quad N_d^F,\quad N_s^F=N_{\bar{s}}^F,\quad N_{\bar{u}}^F=N_{\bar{d}}^F,\nonumber\\
&&\alpha^F_{u,d,s,\bar{u},\bar{d},\bar{s}}\equiv\alpha^F,\quad
\beta^F_{u,d,s,\bar{u},\bar{d},\bar{s}}\equiv\beta^F.
\label{fit1}
\eea

\vspace{0.3cm}

\noindent
FIT 2: Only SGP contribution with seven free parameters: 
\bea
&&N_u^G,\quad N_d^G,\quad N_s^G=N_{\bar{s}}^G,\quad N_{\bar{u}}^G=N_{\bar{d}}^G,\nonumber\\
&&\alpha^G_{u,d,s,\bar{u},\bar{d},\bar{s}}\equiv\alpha^G,\quad \beta^G_{u,d}\equiv\beta^G_v,\quad
\beta^G_{\bar{u},\bar{d},s,\bar{s}}\equiv\beta^G_s.
\label{fit2}
\eea

\vspace{0.3cm}

\noindent
FIT 3: Only SGP contribution with 12 free parameters: 
\bea
&&N_u^G,\quad N_d^G,\quad N_s^G,\quad N_{\bar{u}}^G,\quad N_{\bar{d}}^G,\quad N_{\bar{s}}^G,\nonumber\\
&&\alpha^G_{u},\quad\alpha^G_{d},\quad
\alpha^G_{s,\bar{u},\bar{d},\bar{s}}\equiv\alpha^G_s,
\quad \beta^G_{u},\quad\beta^G_{d},\quad
\beta^G_{\bar{u},\bar{d},s,\bar{s}}\equiv\beta^G_s.
\label{fit3}
\eea

\vspace{0.5cm}

\noindent
Several comments are in order for the above constraints among the parameters: 
With the present limitation of the accumulated SSA data, 
we could not obtain a convergent set of parameters when we treated 
all $N_a^{G,F}$, $\alpha_a^{G,F}$ and $\beta_a^{F,G}$ as independent
in the fit with SGP+SFP contributions.  We thus introduced the above constraint relation 
in FIT 1.  By allowing reasonable different values for $N_a^{G,F}$,
we expect to be able to assume $\alpha_a^{G,F}$ and $\beta_a^{G,F}$ are 
relatively flavor-independent.  
This is particularly so for $\alpha_a^{G,F}$, 
since $A_N$ rises in the forward direction and is expected not to be
very sensitive to $\alpha_a^{F,G}$.   On the other hand, SGP
appears as derivative contribution and thus we assume different values for
$\beta_a^{G}$ between the valence flavors ($a=u,\,d$) and 
the sea flavors ($a=\bar{u},\,\bar{d},\,s,\,\bar{s}$).  
As for the relations among $N_a^{F,G}$ for the ``sea" flavors, we have tried 
all possible combinations among 
$N_{\bar{u}}^{G,F}=\pm N_{\bar{d}}^{G,F}$ and
$N_s^{G,F}=\pm N_{\bar{s}}^{G,F}$.  It turned out that the above
constraint relation in FIT 1 gives the best fit.  
\begin{table}[t]
 \begin{center}
  \begin{tabular}{llll}
   \hline
   \hline
   \multicolumn{4}{c}{FIT 1 : $\chi^2=61.47$}\\
   \hline
   $N_u^G=-0.433$ & $N_d^G=2.38$
   &$N_s^G=N_{\bar{s}}^G=-21.6$ & $N_{\bar{u}}^G=N_{\bar{d}}^G=-19.4$ \\
   $\alpha^G=2.14$ & $\beta^G_{v}=0.682$ &$\beta^G_{s}=-0.956$ \\
   $N_u^F=1.48$ & $N_d^F=-2.98$ & $N_s^F=N_{\bar{s}}^F=-46.7$ &
   $N_{\bar{u}}^F=N_{\bar{d}}^F=6.76$ \\
   $\alpha^F =1.40$ & $\beta^F =3.52$ &  \\
   \hline
   \hline
   \multicolumn{4}{c}{FIT 2 : $\chi^2=132.9$}\\
   \hline
   $N_u^G=-2.05$ & $N_d^G=5.84$ & $N_s^G=N_{\bar{s}}^G=-42.0$
   & $N_{\bar{u}}^G=N_{\bar{d}}^G=-53.8$ \\
   $\alpha^G=2.59$ & $\beta^G_{v}=1.43$ &$\beta^G_{s}=0.932$  \\
   \hline
   \hline
   \multicolumn{4}{c}{FIT 3 : $\chi^2=64.90$}\\
   \hline
   $N_u^G=-2.04\times 10^6$ & $N_d^G=0.237$ & $N_s^G=-2.71$ \\
   $N_{\bar{u}}^G=0.443$ & $N_{\bar{d}}^G=-4.93$ & $N_{\bar{s}}^G=-3.25$ \\
   $\alpha^G_u=4.57$ & $\alpha^G_d=0.178$ & $\alpha^G_s=0.641$ \\
   $\beta^G_u=35.4$ & $ \beta^G_d=-0.180$ & $\beta^G_s=-2.66$  \\
   \hline
   \hline
  \end{tabular}
  \caption{Result for the fitting parameters for FIT 1 (SGP+SFP: 13 parameters), FIT 2
  (SGP only: Seven parameters) and FIT 3 (SGP only: 12 parameters).}
 \end{center}
\end{table}

We have also performed FIT 2 and FIT 3 with only the SGP contribution
in order to see the role of the SFP contribution.
FIT 2 assumes the same relation as the SGP contribution in FIT 1 but
the SFP contribution is omitted, i.e. FIT 2
has less degrees of freedom compared with FIT 1 by the lack of the SFP contribution.  
In FIT 3 with the SGP contribution, we allowed more freedom in the parameters.
This fitting contains the Fit II of \cite{Kouvaris} (hereafter, simply referred to
as KQVY06)
as a special case as well as the above FIT 2.
With this maximally allowed freedom in the SGP contribution, 
the number of degrees of freedom in FIT 3 is almost the same as FIT 1 and thus  
one can see whether the SFP contribution can be replaced by an appropriate
choice of the SGP contribution.  

\begin{figure}[p]
 \large{\underline{STAR $\sqrt{s}=200$ GeV}}
%
 \begin{center}
  \scalebox{0.7}{\includegraphics{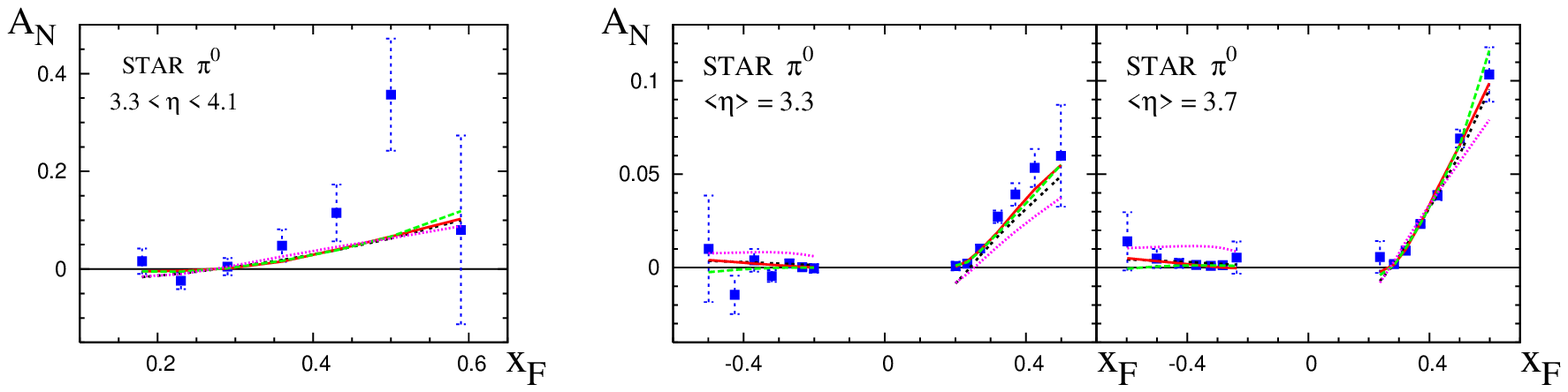}}
 \end{center}
 \large{\underline{BRAHMS $\sqrt{s}=62.4$ GeV}}
\vspace{5mm}
\begin{center}
\hspace{10mm}
\scalebox{0.75}{\includegraphics{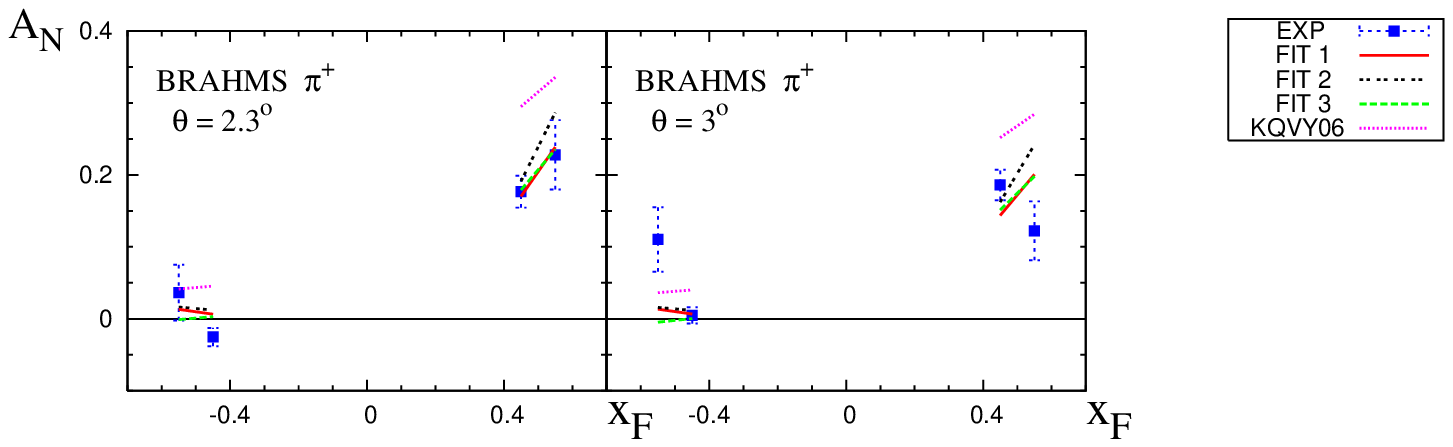}}

\vspace{3mm}
\noindent
\scalebox{0.75}{\includegraphics{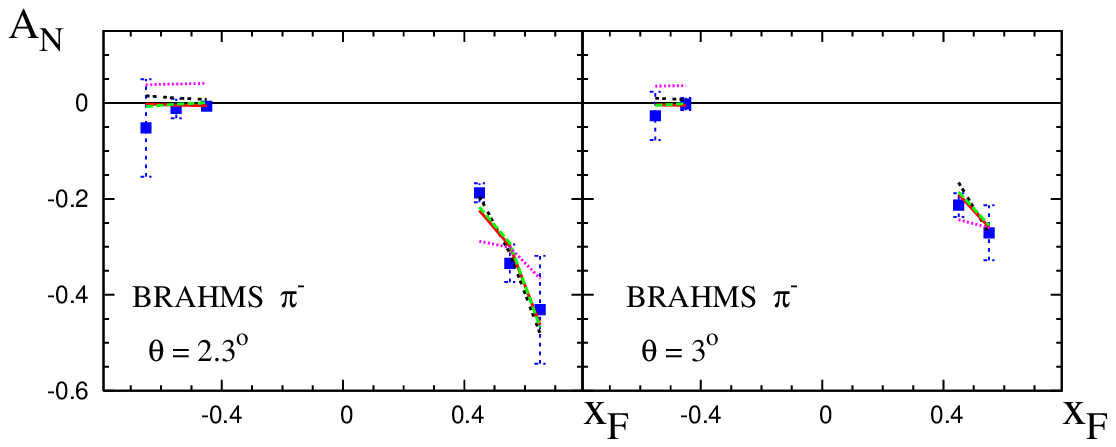}}  

\vspace{3mm}
\noindent
\scalebox{0.75}{\includegraphics{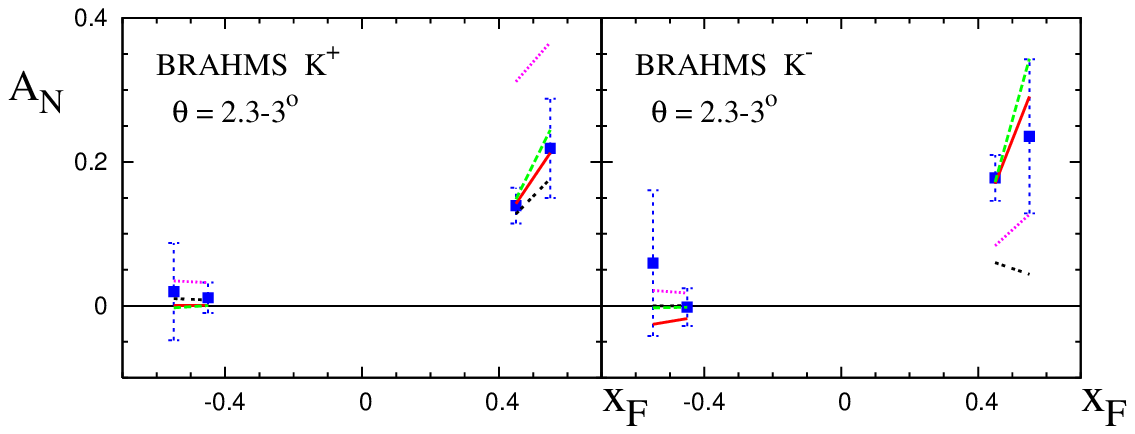}}
 \end{center}
 \caption{The results of the three fits for $A_N$.  The solid lines
are for FIT 1, the long-dotted lines are for FIT 2,
and the dashed lines are for FIT 3.  The result by KQVY06 is also shown  
by the short-dotted lines.}
\end{figure}

The results of the fitting in the above three cases are shown 
in Table 1.   The calculated  $A_N$ are also shown in Fig. 1
together with the RHIC data.  For comparison, we have also shown in the same figure
the result of KQVY06\,\cite{Kouvaris} obtained with
the CTEQ5L distribution function\,\cite{cteq} and the Kretzer's fragmentation function\,\cite{Kretzer}.  
From Fig. 1, one sees that FIT 1 and FIT 3 reproduce $A_N$ data
for all mesons, while FIT 2 and KQVY06 fails for $K^-$. 
We note that our FIT 3 has converged into a result that is different from 
KQVY06. 
This is because we have used different sets of data from KQVY06.  
The deviation of the curves of KQVY06 from data for the charged mesons
is also due to the difference of the used data sets.  
(Also note the kinematics of \cite{BRAHMS}
differs from Fig. 10 of \cite{Kouvaris}.)
This feature is also represented in the resulting $\chi^2$,
which are 61.47, 132.9 and 64.90 for FIT 1, FIT 2, and FIT 3,
respectively.  
Although FIT 3 reproduces $A_N$ equally well as FIT 1,
the behavior of the obtained fitting functions are extreme and 
looks unphysical as is seen by the resulting parameters in the functions
shown in Table 1.  
This suggests that the SFP contribution is also a necessary source for SSA
and plays an important role which cannot be substituted by the SGP
contribution.

\begin{figure}
 \begin{minipage}{0.3\hsize}
  \begin{center}
   \scalebox{0.6}{\includegraphics{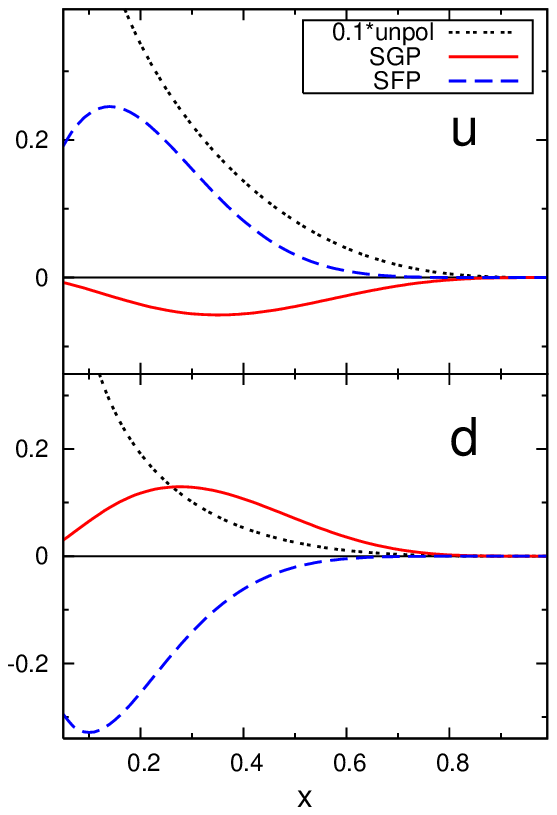}}
  \end{center}
 \end{minipage}
 \begin{minipage}{0.3\hsize}
 \begin{center}
  \scalebox{0.6}{\includegraphics{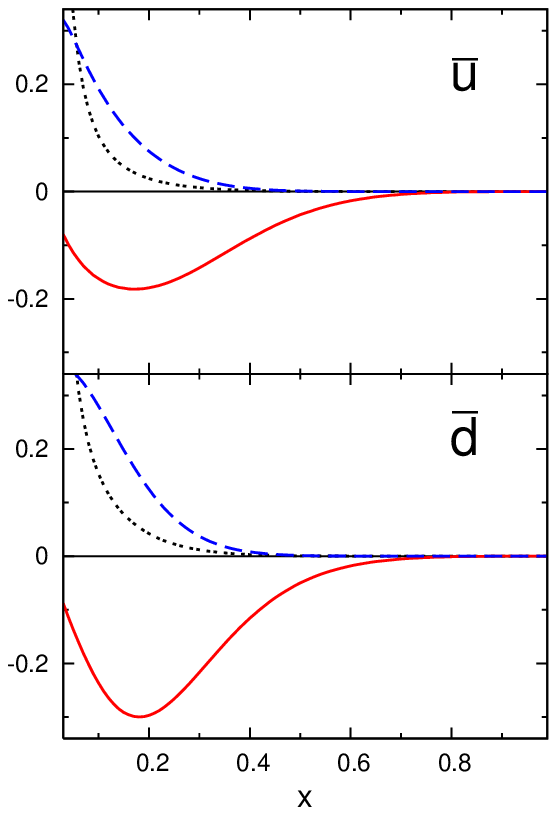}}
 \end{center}
 \end{minipage}
 \begin{minipage}{0.3\hsize}
 \begin{center}
  \scalebox{0.6}{\includegraphics{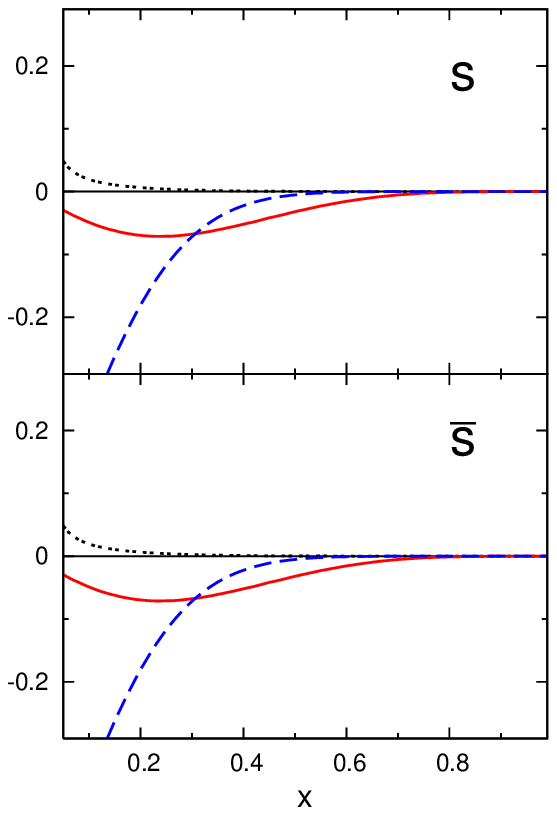}}
 \end{center}
 \end{minipage}
 \caption{The SGP function $G_{F}^{a}(x,x)$
and the SFP function $G_{F}^{a}(0,x)+\widetilde{G}_{F}^{a}(0,x)$
at $Q^2=2.4$ {GeV}$^2$
for each quark-flavor obtained in FIT 1 in comparison to
the unpolarized quark distribution $f_a(x)$ [scaled by factor 1/10].}
\end{figure}

\begin{figure}
 \begin{minipage}{0.3\hsize}
  \begin{center}
   \scalebox{0.85}{\includegraphics{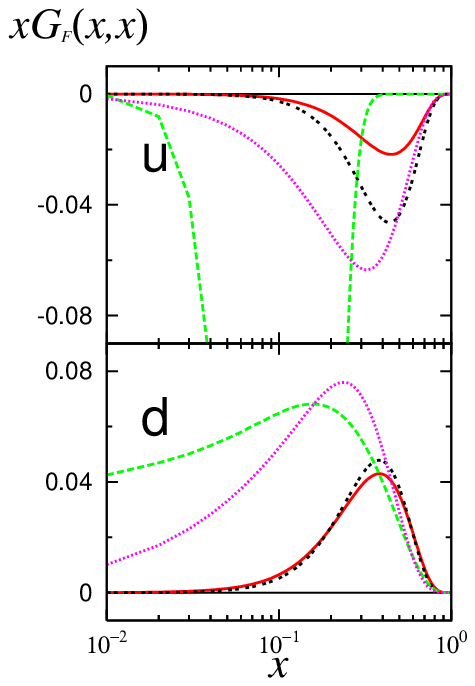}}
  \end{center}
 \end{minipage}
 \begin{minipage}{0.3\hsize}
 \begin{center}
  \scalebox{0.85}{\includegraphics{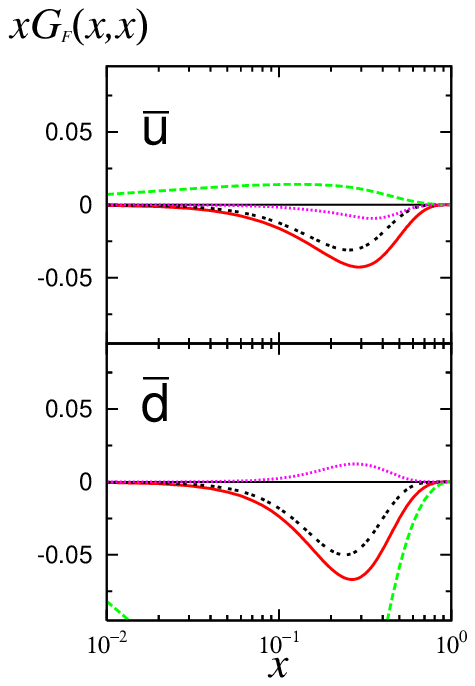}}
 \end{center}
 \end{minipage}
 \begin{minipage}{0.3\hsize}
 \begin{center}
  \scalebox{0.85}{\includegraphics{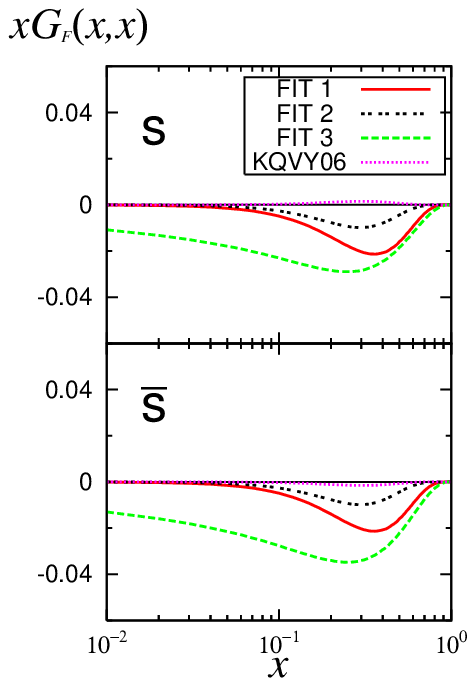}}
 \end{center}
 \end{minipage}
 \caption{SGP functions at $Q^2=2.4$ {GeV}$^2$ obtained in three fits.  
The SGP function in KQVY06 is
also shown by using the relation $G_{F}^{a}(x,x)=-\frac{\sqrt{4\pi\alpha_s}}{\pi M_N}
 T_{F,a}(x,x)$ for each quark flavor. For antiquark flavors, the sign of this
 relation is opposite. }
\end{figure}

Figure 2 shows the obtained SGP and SFP functions in FIT 1 for each
quark flavor
in comparison with the unpolarized parton density (scaled by factor 1/10).  
One sees that the SGP and SFP functions are comparable in magnitude,
but the SFP functions are more shifted in smaller $x$ regions.  
Among SGP functions, those for the ``valence" flavors ($u$ and $d$ quarks)
spread more into a larger $x$ region than those for sea flavors.
This spreading into a larger $x$ region actually plays an important role
for the rising $A_N$ in the forward region.  
We observe that
the flavor dependence of the SFP functions also has a similar tendency.   
Owing to the behaviors of each function, the largest contribution to
$A_N^{\pi^+}$ and $A_N^{\pi^-}$ are, respectively, from $u$ and $d$ quark SGP functions.  
As for the kaons, the SGP contribution from strange quark turned out to be the largest
with small corrections from other flavors and the SFP contributions.
This is largely due to the large strangeness component in the
DSS fragmentation function.  
From the definition of the SGP and SFP functions, 
the latter might be interpreted as a more ``exotic" component of the
correlations, while the former may be closer to a quark parton density accompanied by the
soft gluon.  This may lead to the SFP function to have nonzero values only in the smaller
$x$ region.  

To compare the functions in FIT 1 with other cases,
we have shown in Fig. 3 the SGP functions obtained in each fit.
For comparison we have also shown the SGP functions of KQVY06.    
As is seen from Fig. 3, the SGP functions for $u$ and $\bar{d}$ quarks in FIT 3
are huge compared with those for other quark-flavors, and the former
is sharply peaked, which are the result 
of the extreme values of the parameters in FIT 3 shown in Table 1.  
This unphysical behavior of the SGP function can be taken as a manifestation 
of the fact that the SGP function alone can not reproduce all of the observed $A_N$,
and other sources of SSA are necessary.  
From Fig. 3,
the SGP functions in FIT 1 turn out to be smaller compared with those
in FIT 2 and FIT 3, since the SFP functions
can give rise to the additional $A_N$s.  
Moderate behaviors of the SGP and SFP functions given in FIT 1 as shown in 
Figs. 2 and 3 indicate 
that the SFP contribution is a natural source of $A_N$
together with the SGP contribution.  

\begin{figure}[p]
 \large{\underline{STAR $\sqrt{s}=200$ GeV}}
%
 \begin{center}
  \scalebox{0.7}{\includegraphics{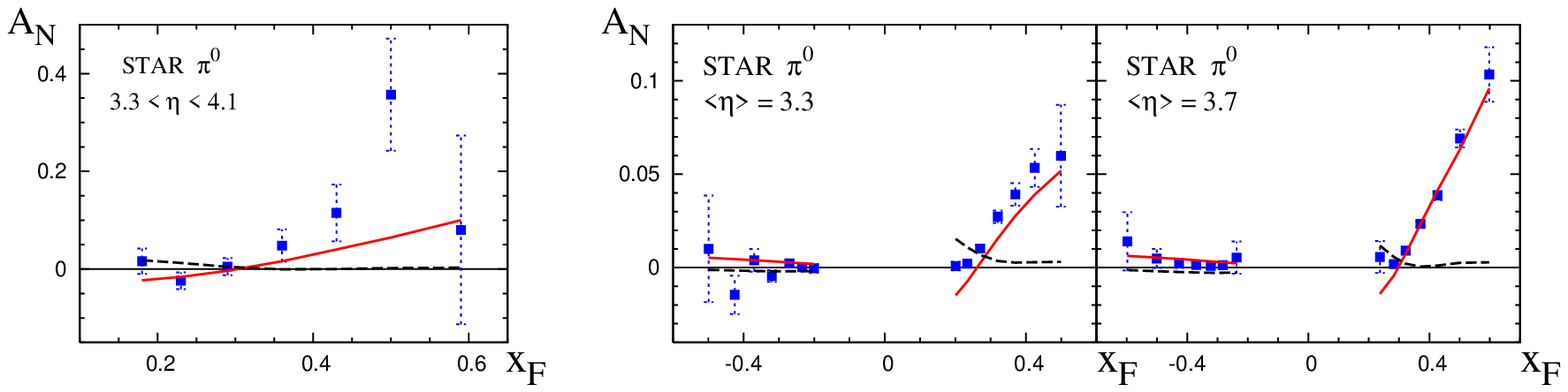}}
 \end{center}
 \large{\underline{BRAHMS $\sqrt{s}=62.4$ GeV}}
\vspace{5mm}
\begin{center}
\scalebox{0.75}{\includegraphics{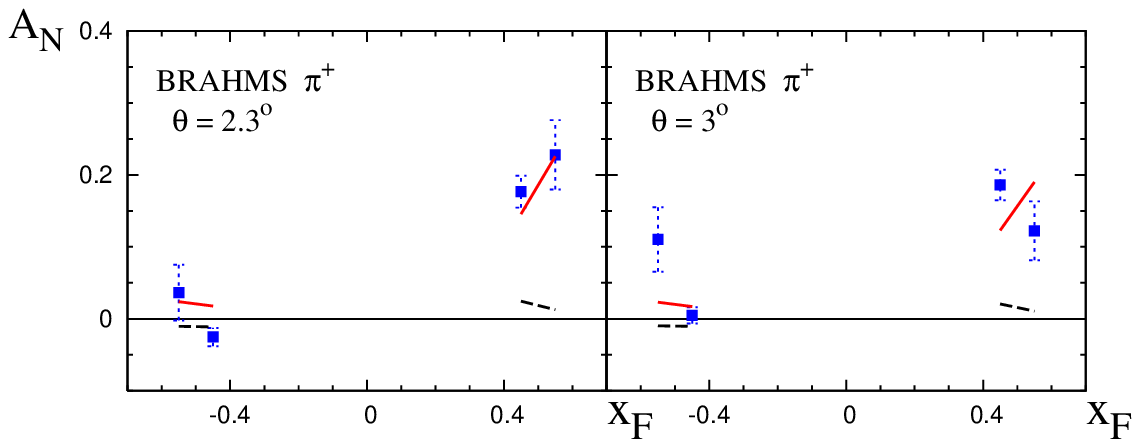}}

\vspace{3mm}
\noindent
\scalebox{0.75}{\includegraphics{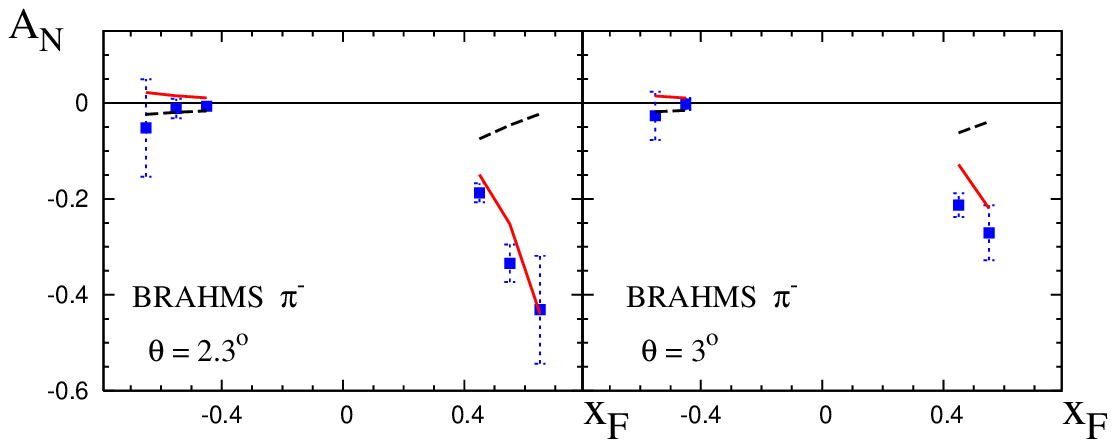}}  

\vspace{3mm}
\noindent
\scalebox{0.75}{\includegraphics{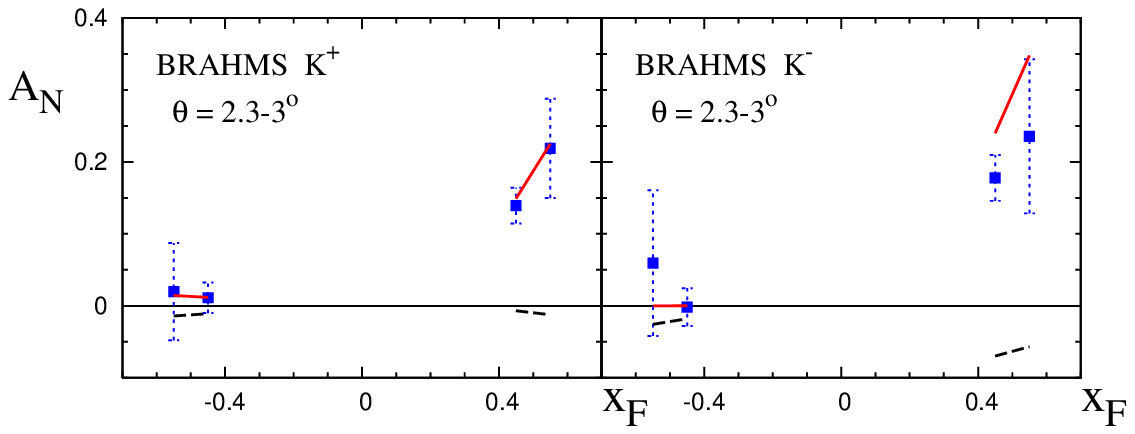}}
 \end{center}
 \caption{Separation of $A_N$ in FIT 1 
into the SGP (solid lines) and the SFP (dashed lines) contributions. }
\end{figure}

To see the relative magnitude of the SFP contributions in FIT 1, we have shown in Fig. 4 
the decomposition of $A_N$ into the SGP and SFP contributions.
One sees from the figure that in many cases
the SGP brings a larger contribution as a whole, while the SFP works as a small but non-negligible
correction.
For the kaons, we observed that the SFP contribution from $u$ and $d$
quarks are quite large but they tend to cancel each other.   
As one can see from (\ref{tSGP}) and (\ref{tSFP})
the SFP hard cross sections are larger than the SGP hard cross section
typically by factor $N_c^2$\,\cite{KoikeTomita}. 
However, the SGP functions  
spread more in the larger $x$ region as shown in Fig. 2 and also contribute as a
derivative.  Therefore it brings a large 
contribution to $A_N$ in the forward region.

\begin{figure}
  \scalebox{0.9}{\includegraphics{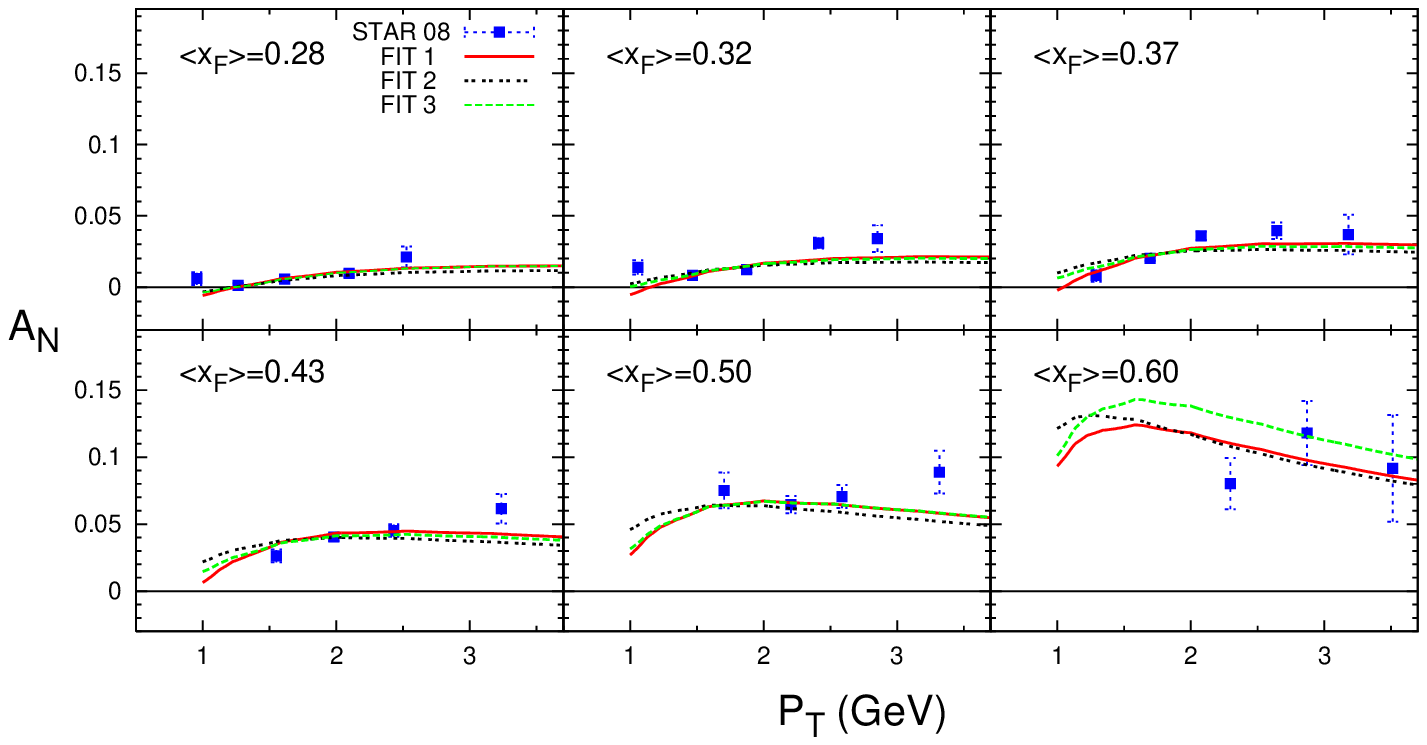}}
 \caption{Comparison of the $P_{hT}$-dependence of $A_N^{\pi^0}$ with the result of
 STAR data\,\cite{STAR2}.}
\end{figure}

Figure 5 shows the $P_{hT}$ dependence of $A_N^{\pi^0}$ for $\sqrt{S}=200 $ GeV
together with the STAR data.  All fits give similar curves 
for each $x_F$ and approximately reproduce experimental data.  
This is quite natural because our three fits  
used correct $P_{hT}$ at each $x_F$ of the $A_N$ data
shown in 
Fig. 1. If we had included E704 data in the fit without
taking into account of the $P_{hT}$ dependence, deviation from the RHIC data
may have been observed in the 
$P_{hT}$-depndence as was the case in the analysis by KQVY06.

\begin{figure}
 \begin{minipage}{0.45\hsize}
  \begin{center}
   \scalebox{0.7}{\includegraphics{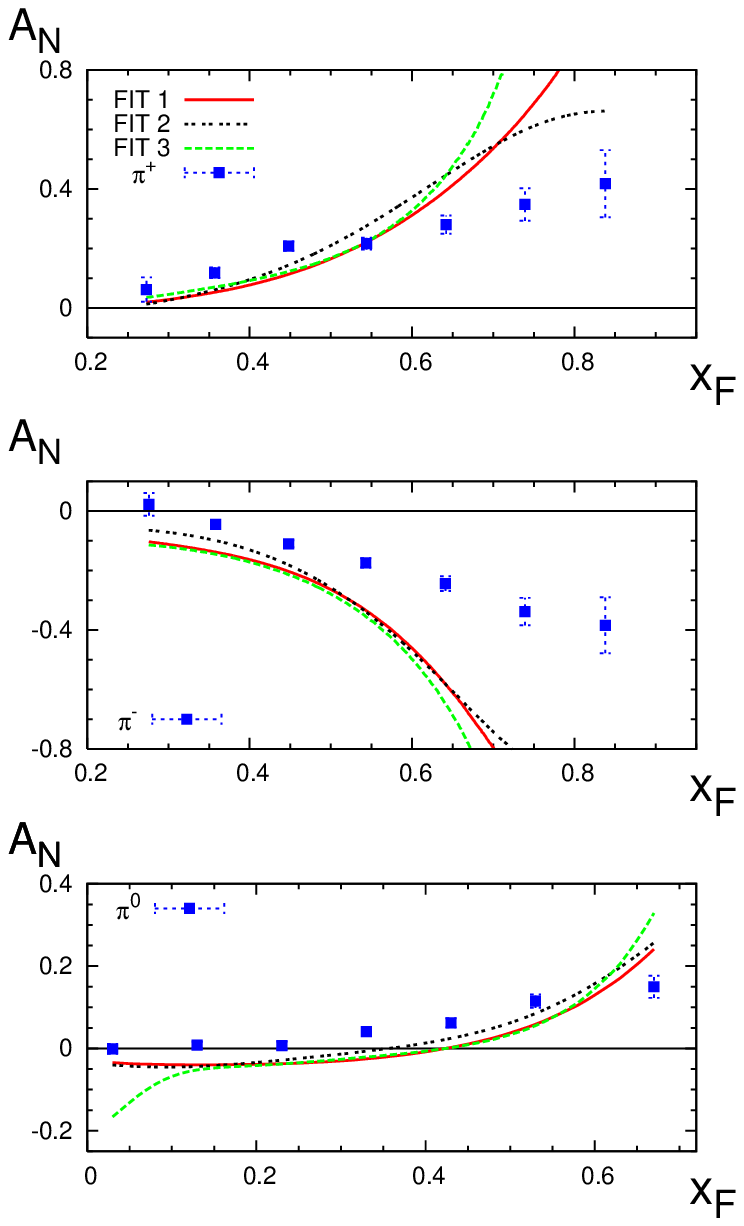}}
  \end{center}
 \end{minipage}
 \begin{minipage}{0.45\hsize}
 \begin{center}
  \scalebox{0.7}{\includegraphics{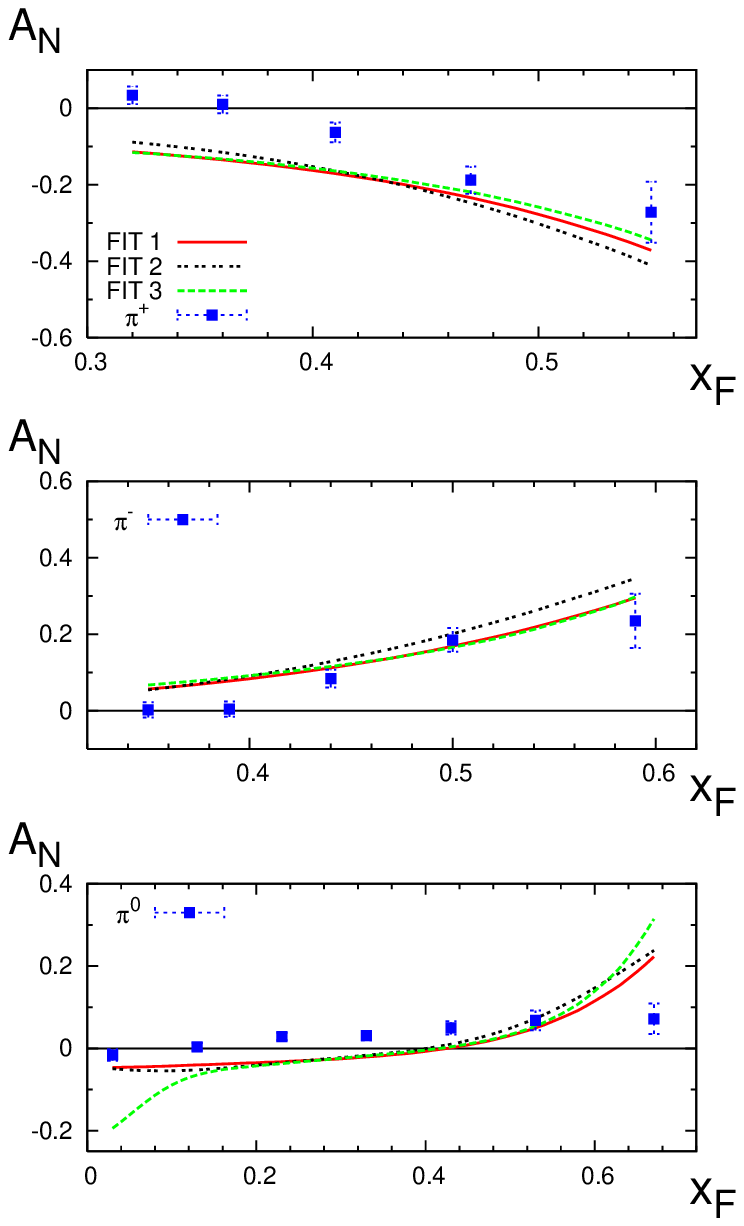}}
 \end{center}
 \end{minipage}
 \caption{Comparison of the fits with the FNAL-E704 data.  We set $P_{hT}=1.2$ (GeV)
in the theory calculation. The left three figures are for the
$pp$ collision, and the right three figures are for the $\bar{p}p$
 collision. }
\end{figure}

Using the SFP and SGP functions in the fits, we have calculated
$A_N^\pi$ at the FNAL energy $\sqrt{S}=20$ $\textrm{GeV}$
for $p^\uparrow p\to\pi X$ and $\bar{p}^\uparrow p\to\pi X$.  
As mentioned before, we did not include the $A_N$ data of FNAL-E704 
($\sqrt{S}=20$ GeV) in the fitting.    
We do this comparison as a semiquantitative test of our fit result in terms of the twist-3 formalism
at a lower energy. 
With our convention for the twist-3 distributions for the ``antiquark" flavor
in (\ref{anti}), those in the antiproton are related to those in the proton as
\bea
G_{F}^{a}(x_1,x_2)|_{\rm anti-proton}=G_{F}^{\bar{a}}(x_1,x_2)|_{\rm proton},\qquad
\widetilde{G}_{F}^{a}(x_1,x_2)|_{\rm anti-proton}=
\widetilde{G}_{F}^{\bar{a}}(x_1,x_2)|_{\rm proton}.  
\label{pbar}
\eea
The result was shown in Fig. 6 together with the E704 data\,\cite{E704}.
From Fig. 6, one sees that FIT 1 and FIT 2 give qualitatively similar
curves as the data,  while FIT 3 shows the deviation from the data at
small-$x_F$ for $\pi^0$.
As pointed out in \cite{hatan}, the NLO QCD in the collinear factorization
can not reproduce the unpolarized cross section at the fixed target energy
and thus our twist-3 formula for $A_N$ may not be directly used for the FNAL data.  
With this in mind, 
KQVY06 included the E704 data in their fitting by multiplying
the ad hoc factor $N_{\rm E704}=1/2$ to the theory calculation.
If we had known this $N_{\rm E704}$ more precisely, 
we could have included the E704 data in our analysis,
which is beyond the scope of the present study.  
From Fig. 6, one sees the E704 data is smaller than the calculated
$A_N$, which is consistent with the analysis in \cite{Kouvaris}.   

\section{Summary}

In this paper, we have presented a numerical analysis of 
the RHIC $A_N$ data for $p^\uparrow p\to hX$ ($h=\pi,\ K$)
in terms of the twist-3 mechanism in the collinear factorization.  
As a source of the asymmetry we have included all the contribution
from the twist-3 quark-gluon correlation functions
in the transversely polarized nucleon, which
consist of the SGP and SFP contributions.  
We have shown that the combination of the SGP and SFP contributions can 
reproduce the $A_N$ data at RHIC
with moderate choice of the functions, while it is difficult to reproduce all of them 
if one keeps only the SGP
contribution.  
In particular, $A_N$ for $K^\pm$ was properly reproduced by the
inclusion of
the SFP contribution.  
Although the relevance of the SFP contribution is
numerically supported, we do not have any dynamical information
on the shape and magnitude of the SGP and SFP functions at this stage.  
\footnote{There is, however, a model calculation
\,\cite{Kang:2010hg} showing the SFP function is small.}
In addition, our present analysis did not include the second term
of (\ref{3tu}), which could be another important source of SSA.  
To clarify these points, one needs more variety of SSA data 
together with
the nonperturbative study on the SGP and SFP functions.  

\vspace{0.5cm}

\noindent
{\bf Acknowledgement}

We thank D. de Florian for providing us with the Fortran code
of their fragmentation function.

\newpage



\begin{thebibliography}{99}

\vspace{0.5cm}

\bibitem{E704}
D.L. Adams {\it et al.} (E704 Collaboration), Phys. Lett. {\bf B261},
	201 (1991); {\bf 264}, 462 (1991). 

\bibitem{COMPASS}
V.Y. Alexakhin {\it et al.}(COMPASS Collaboration), Phys. Rev. Lett {\bf 94}, 202002 (2005);
E.S. Ageev {\it et al.} (COMPASS Collaboration), Nucl. Phys. {\bf B765},
	31 (2007); M. Alekseev {\it et al.} (COMPASS Collaboration), Phys. Lett. {\bf B673}, 127 (2009). 

\bibitem{HERMES}
A. Airapetian {\it et al.} (HERMES Collaboration), Phys. Rev. {\bf D64}, 097101 (2001); Phys. Rev. Lett. {\bf 94}, 012002 (2005); {\bf 103}, 152002 (2009). 


\bibitem{STAR1}
J. Adams {\it et al.} (STAR Collaboration), Phys. Rev. Lett. {\bf 92}, 171801 (2004). 

\bibitem{STAR2}
B.I. Abelev {\it et al.} (STAR Collaboration), Phys. Rev. Lett. {\bf 101}, 222001 (2008). 

\bibitem{PHENIX} S.~S. Adler {\it et al.}
(PHENIX Collaboration), Phys. Rev. Lett. {\bf 95}, 202001 (2005). 

\bibitem{BRAHMS}
I. Arsene {\it et al.} (BRAHMS Collaboration), Phys. Rev. Lett. {\bf 101}, 042001 (2008). 

\bibitem{Kane}
G.L. Kane, J. Pumplin, and W. Repko,
Phys. Rev. Lett. {\bf 41}, 1689 (1978).  


\bibitem{Sivers} D.~W. Sivers, Phys. Rev. {\bf D41}, 83 (1990);
	Phys. Rev. {\bf D43}, 261 (1991).

\bibitem{Collins} J.~C. Collins, Nucl. Phys. {\bf B396}, 161 (1993).

\bibitem{Collins02} J.~C. Collins, Phys. Lett. {\bf B536}, 43 (2002).

\bibitem{BJY03} A.~V. Belitsky, X.~D. Ji, and F. Yuan, Nucl. Phys. {\bf B656}, 165 (2003).

\bibitem{BMP03} D. Boer, P. Mulders and F. Pijlman, Nucl. Phys. {\bf B667}, 201 (2003).

\bibitem{CM04} J.~C. Collins and A. Metz, Phys. Rev. Lett. {\bf 93}, 252001 (2004).

\bibitem{MM09}
S. Meissner and A.Metz,
Phys. Rev. Lett. {\bf 102}, 172003 (2009). 

\bibitem{CS81} J.C. Collins and D.E. Soper, Nucl. Phys. {\bf B193}, 381 (1981);
{\bf B213}, 545(E) (1983). 

\bibitem{CSS85} J.C. Collins, D.E. Soper, and G. Sterman, Nucl. Phys. {\bf B250}, 199 (1985).

\bibitem{JMY05} X.~D. Ji, J.~P. Ma, and F. Yuan, Phys. Rev. {\bf D71}, 034005 (2005); 
Phys. Lett. {\bf B597}, 299 (2004).


\bibitem{BMP04} 
C.~J. Bomhof, P.~J. Mulders, and F. Pijlman, Phys. Lett. {\bf B596}, 277 (2004);
Eur. Phys. J. {\bf C47}, 147 (2006); \\ 
A. Bacchetta, C.~J. Bomhof, P.~J. Mulders, and F. Pijlman, 
Phys. Rev. {\bf D72}, 034030 (2005);\\
C.J. Bomhof and P.J. Mulders, JHEP 0702:029, (2007);\\
C.J. Bomhof and P.J. Mulders, Nucl. Phys. {\bf B795}, 409 (2008). 



\bibitem{Anselmino}
M. Anselmino, M. Boglione, U. D'Alesio, A. Kotzinian, F. Murgia, and
	A. Prokudin, Phys. Rev. {\bf D71}, 074006 (2005); \\
M. Anselmino, M. Boglione, U. D'Alesio, A. Kotzinian, F. Murgia,
	A. Prokudin, and C.T\"urk, Phys. Rev. {\bf D75}, 054032 (2007); 	\\
M. Anselmino, M. Boglione, U. D'Alesio, A. Kotzinian, S. Melis, F. Murgia,
 A. Prokudin, and C. T\"urk, 
Eur. Phys. J. {\bf A39}, 89 (2009);\\
and references quoted therein. 


\bibitem{ET}
A.V. Efremov and O.V. Teryaev, Sov. J. Nucl. Phys. {\bf 36}, 140 (1982),
	Yad. Fiz. {\bf 36}, 242 (1982);

A.V. Efremov and O.V. Teryaev, Phys. Lett. {\bf B150}, 383 (1985). 

\bibitem{QS92}
J.W. Qiu and G. Sterman,
Nucl. Phys. {\bf B378}, 52 (1992). 

\bibitem{QS99}
J.W. Qiu and G. Sterman,
Phys. Rev. {\bf D59}, 014004 (1998). 

\bibitem{Kanazawa}
Y. Kanazawa and Y. Koike, Phys. Lett. {\bf B478}, 121 (2000);

Y. Kanazawa and Y. Koike, Phys. Lett. {\bf B490}, 99 (2000). 

\bibitem{EKT}
H. Eguchi, Y. Koike, and K. Tanaka, 
Nucl. Phys. {\bf B752}, 1 (2006); 

H. Eguchi, Y. Koike, and K. Tanaka, 
Nucl. Phys. {\bf B763}, 198 (2007). 

\bibitem{Kouvaris}
C. Kouvaris, J.W. Qiu, W. Vogelsang, and F. Yuan, 
Phys. Rev. {\bf D74}, 114013 (2006). 

\bibitem{KT071}
  Y.~Koike and K.~Tanaka,
  Phys.\ Lett.\  B {\bf 646}, 232 (2007)
  [Erratum-ibid.\  B {\bf 668}, 458 (2008)]

\bibitem{KT072}
Y. Koike and K. Tanaka,
Phys. Rev. {\bf D76}, 011502(R) (2007).  

\bibitem{KQ08}
  Z.~B.~Kang and J.~W.~Qiu,
  Phys.\ Rev.\  D {\bf 78}, 034005 (2008).\\
  Z.~B.~Kang, J.~W.~Qiu, W.~Vogelsang, and F.~Yuan, Phys. Rev. {\bf D78} 114013 (2008). 

\bibitem{KoikeTomita}
Y. Koike and T. Tomita, 
Phys. Lett. {\bf B675}, 181 (2009). 


\bibitem{OL}
X.D. Ji, J.W. Qiu, W. Vogelsang, and F. Yuan, Phys. Lett. {\bf 97}, 082002 (2006);

X.D. Ji, J.W. Qiu, W. Vogelsang, and F. Yuan, Phys. Rev. {\bf D73}, 094017 (2006);
 
X.D. Ji, J.W. Qiu, W. Vogelsang, and F.Yuan, Phys. Lett. {\bf B638}, 178 (2006);

Y. Koike, W. Vogelsang, and F. Yuan, Phys. Lett. {\bf B659}, 878 (2008).


\bibitem{YZ09}
F. Yuan and J. Zhou, Phys. Rev. Lett. {\bf 103}, 052001 (2009);\\
Z. Kang, F. Yuan, and J. Zhou,
arXiv:1002.0399:[hep-ph]. 

\bibitem{unpol}
B.L. Combridge, J. Kripfganz, and J. Ranft, Phys. Lett. {\bf B70}, 234 (1977);\\
R. Cutler and D.W. Sivers, Phys. Rev. {\bf D17}, 196 (1978);\\
J.F. Owens, E. Reya, and M. Gl\"uck, Phys Rev. {\bf D18}, 1501 (1978). 


\bibitem{hatan}
C. Bourrely and J. Soffer,
Eur. Phys. J. {\bf C36}, 371 (2004). 


\bibitem{Braun}
  V.~M.~Braun, A.~N.~Manashov, and B.~Pirnay,
  Phys.\ Rev.\  D {\bf 80}, 114002 (2009).


\bibitem{GRV98}
M. Gl\"uck, E. Reya, and A. Vogt,
Eur. Phys. J. {\bf C5}, 461 (1998)

\bibitem{DSS07}
D. de Florian, R. Sassot, and M. Stratmann, Phys. Rev. {\bf D75}, 114010 (2007). 

\bibitem{cteq}
H.L. Lai {\it et al.} (CTEQ Collaboration), Eur. Phys. J. {\bf C12}, 375
	 (2000). 

\bibitem{Kretzer}
S. Kretzer,
Phys. Rev. {\bf D62}, 054001 (2000).

 
\bibitem{Kang:2010hg} Z.~B.~Kang,
    J.~W.~Qiu and H.~Zhang, 
arXiv:1004.4183 [hep-ph].



\end{thebibliography}
\end{document}